\documentstyle[12pt,epsf]{article}
\def\href#1#2{#2}   
\newif\ifdraft
\draftfalse	  
\reversemarginpar   
\makeatletter
\let\mlabel=\label
\let\adkendequation=\endequation%
\def\endequation{\adkendequation\adklabel\global\@ignoretrue}
\let\adkendeqnarray=\endeqnarray%
\def\endeqnarray{\adkendeqnarray\adklabel\global\@ignoretrue}
\newbox\marglabbox
\def\adklabel{\ifvoid\marglabbox\else\marginpar{\unhbox\marglabbox}\fi}
\def\label#1{\ifdraft\ifmmode%
  \global\setbox\marglabbox=\hbox{\hfill\fbox{\tiny\verb*~#1~}}%
  \else\ifinner\else\marginpar{\hfill\fbox{\tiny\verb*~#1~}}%
  \fi\fi\fi \mlabel{#1}}
\makeatother
\ifdraft%
\fi
\font\twelvebb=msbm12
\font\tenbb=msbm10
\font\sevenbb=msbm7
  \newfam\bbfam
  
  \textfont\bbfam=\twelvebb
  \scriptfont\bbfam=\tenbb
  \scriptscriptfont\bbfam=\sevenbb
\font\twelveeusm=eusm10 scaled 1200
\font\teneusm=eusm10
  \newfam\eusmfam
  
  \textfont\eusmfam=\twelveeusm
  \scriptfont\eusmfam=\teneusm
  \scriptscriptfont\eusmfam=\scriptfont\eusmfam
\font\twelvefrak=eufm10 scaled 1200
\font\tenfrak=eufm10
  \newfam\frakfam
  
  \textfont\frakfam=\twelvefrak
  \scriptfont\frakfam=\tenfrak
  \scriptscriptfont\frakfam=\scriptfont\frakfam
\def\sqr#1#2{{\vcenter{\hrule height.#2pt
   \hbox{\vrule width.#2pt height#1pt \kern#1pt
      \vrule width.#2pt}
   \hrule height.#2pt}}}

\def\bsqr#1#2{{\vrule width #1pt height#2pt}}
\def\bsquare{{\mathchoice\bsqr66\bsqr66\bsqr33\bsqr33}}
\def\badbreak{\penalty1000}

\newcommand{\cE}{{\cal E}}                  
\newcommand{\cS}{{\cal S}}                  
\newcommand{\lsz}{{\ell}}                   
\newcommand{\cml}{{\cal C}}                 

\begin{document}

\begin{center}
{\Large{\bf Inherently Global Nature of Topological}} \\
\vspace*{.1in}
{\Large{\bf Charge Fluctuations in QCD}}\\
\vspace*{.4in}
{\large{I.~Horv\'ath$^1$,
A.~Alexandru$^1$,
J.B.~Zhang$^2$,
Y.~Chen$^3$,
S.J.~Dong$^1$,
T.~Draper$^1$}}\\
\vspace*{.1in}
{\large{
F.X.~Lee$^4$,
K.F.~Liu$^1$,
N.~Mathur$^1$,
S.~Tamhankar$^1$ and
H.B.~Thacker$^6$}} \\
\vspace*{.15in}
$^1$University of Kentucky, Lexington, KY 40506\\
$^2$CSSM and Department of Physics and Mathematical Physics, 
University of Adelaide, Adelaide, SA 5005, Australia\\
$^3$Institute of High Energy Physics, Academia Sinica, Beijing 100039, P.R. China\\
$^4$Center for Nuclear Studies and Department of Physics, 
George Washington University, Washington, DC 20052 \\
$^6$Department of Physics, University of Virginia, Charlottesville, VA 22901

\vspace*{0.2in}
{\large{Jan 16 2005}}

\end{center}

\vspace*{0.15in}

\begin{abstract}
  \noindent
  We have recently presented evidence that in configurations dominating the regularized 
  pure-glue QCD path integral, the topological charge density constructed from the overlap 
  Dirac operator organizes into an ordered space-time structure. It was pointed out 
  that, among other properties, this structure exhibits two important features: it is 
  {\em low-dimensional} and {\em geometrically global}, i.e. consisting of connected 
  sign-coherent regions with local dimensions $1\le d < 4$, and spreading over 
  arbitrarily large space--time distances. Here we show that the space-time structure 
  responsible for the origin of topological susceptibility indeed exhibits global behavior. 
  In particular, we show numerically that topological 
  fluctuations are not saturated by localized concentrations of most intense topological 
  charge density. To the contrary, the susceptibility saturates only after the space-time 
  regions with most intense fields are included, such that geometrically global structure 
  is already formed. We demonstrate 
  this result both at the fundamental level (full topological density) and at low energy 
  (effective density). The drastic mismatch between the point of fluctuation saturation 
  ($\approx$ 50\% of space-time at low energy) and that of global structure formation 
  ($<$ 4\% of space-time at low energy) indicates that the ordered space-time structure in 
  topological charge is {\em inherently global} and that topological charge fluctuations 
  in QCD {\em cannot} be understood in terms of individual localized pieces. Description 
  in terms of global brane-like objects should be sought instead.
\end{abstract}

\noindent
{\bf 1. Introduction.} 
Understanding the nature of topological charge fluctuations is an essential step toward 
clarifying the structure of the QCD vacuum. Indeed, not only are there important physical
phenomena associated with topological charge (such as $\eta'$ mass, $\theta$--dependence
and possibly spontaneous chiral symmetry breaking (SChSB)), but there are also intriguing 
theoretical reasons to believe that topological charge structure in QCD may provide 
a particularly clear window on the gauge theory/string theory connection, since topological 
charge is dual to Ramond-Ramond charge in type IIA string theory~\cite{Witten98}. 
Nevertheless, it is only recently that a first-principles {\em direct} study of space-time 
topological structure in a well-defined lattice-regularized setting was carried 
out~\cite{Hor03A,Hor02B}. 
This development was made possible by important advances in understanding lattice chiral 
symmetry~\cite{chiral,NarNeu95,Neu98BA} which, among other things, led to the construction 
of novel lattice topological charge density operators~\cite{Has98A,NarNeu95}. These operators 
are not just convenient in that the corresponding topological susceptibility can be computed 
without any subjective manipulations of the gauge field and that theoretical properties of 
the lattice topological field match those in the continuum~\cite{topo_th}. In addition, they 
exhibit features that are inherent to smooth backgrounds on continuum manifolds, but quite 
unexpected for general backgrounds contributing to the regularized QCD path integral. 
In particular, the index theorem is exactly satisfied relative to the lattice Dirac operator 
used to define the density~\cite{Has98A}, and the global charge of a generic configuration
is strictly stable with respect to small local changes of the gauge field.

In previous work we have studied the space-time behavior of such topological density based
on the overlap Dirac operator in pure-glue QCD~\cite{Hor03A}. We found a clear excess of 
ordered structure compared to random distributions of charge. In fact, topological charge 
organizes into two oppositely-charged {\em low-dimensional ``sheets''} that fill a macroscopic
fraction of space-time, and are built around a connected {\em ``skeleton''} of points with 
high intensity~\footnote{From the geometric point of view, both the sheets and the skeleton 
behave somewhat similarly to Peano's space-filling curve~\cite{Hor03A}.}. This structure is 
embedded in space-time in a non-trivial geometric manner and it should be regarded as fundamental 
since it includes fluctuations at all scales up to the lattice cutoff.~\footnote{Analogous 
structure was recently reported to be seen also in the 2-d CP(N-1) models~\cite{Tha04}.} 
We note that a viable candidate for the fundamental structure in typical configurations 
contributing to the QCD path integral must reproduce the correct behavior of the topological 
density 
correlator~\cite{Hor02B}. A particularly notable requirement is that it has to conform to the fact 
that $\langle q(0) q(x) \rangle \le 0$ at arbitrary non-zero distance~\cite{SeSt}. Indeed, for 
the above two-sheet/skeleton structure the non-positivity of the correlator at non-zero distances 
arises in an ordered manner due to the fact that layers of low-dimensional structure with 
alternating sign permeate the space-time.

Two basic aspects of the topological charge structure described in~\cite{Hor03A} are 
particularly intriguing and possibly far-reaching. First is the strictly {\em low-dimensional} 
nature of the structure~\footnote{The low-dimensional nature of fundamental topological field
is reflected to some degree also in low-lying Dirac modes~\cite{Hor02A,Aub04}. 
More precise form of this correspondence is yet to be understood. The notion of strictly 
low-dimensional structure is also emerging using indirect projection techniques~\cite{Zakh}.
Its relation (if any) to the structure in topological field is yet to be understood.}.
By that we mean that the structure only contains parts with 
local dimensions strictly less than four~\footnote{The dimension is at least one since 
the sign-coherent sheet/skeleton is a connected global structure.}. This suggests some 
interesting possibilities for the way in which gauge fields guide the propagation of pions, 
and will hopefully shed new insight into the relation between topological charge fluctuations 
and SChSB. The second aspect is the {\em geometrically global} nature of the structure. This 
expresses a geometric fact that the subset of space-time occupied by the structure (sign-coherent 
sheet or skeleton) is a connected set spreading over maximal available distances. 
In this paper we focus on the distribution of charge within the geometric structure and 
show that the global behavior is {\em necessarily present} in the physically relevant 
substructure responsible for the origin of topological susceptibility. In particular, 
we demonstrate that the bulk of topological susceptibility in QCD cannot be carried 
by localized concentrations of most intense topological charge. Just to the contrary, we will 
argue quantitatively that the space-time structure carrying topological charge fluctuations 
is {\em inherently global (super-long-distance)}, in the sense discussed in Ref.~\cite{Hor04A}. 
In other words, our results indicate that this structure (both at the fundamental level and at 
low energy) cannot be broken into localized objects that could be used as a basis for valid 
physical analysis of the origin of topological susceptibility in QCD. 
Analysis in terms of global low-dimensional brane-like objects should be sought instead.

\medskip

\noindent
{\bf 2. The Argument.} 
The methodology for analysis of global behavior in space-time structure of local observables
has been developed in Ref.~\cite{Hor04A}. Let us first review the basic ingredients of this
approach in case of topological charge density. In what follows, we will assume that $q(x)$
is a well-defined lattice topological density in the sense that the topological
susceptibility can be directly computed via the associated correlator or, equivalently,
via $\chi = (\langle Q^2 \rangle - \langle Q \rangle^2)/V$, where $Q=\sum_x q(x)$. 
If $D$ is the overlap Dirac operator~\cite{Neu98BA} based on the Wilson-Dirac kernel
with mass $-\rho$, then the density~\cite{Has98A}
\begin{equation}
   q(x) \;=\; \frac{1}{2\rho} \, \mbox{\rm tr} \,\gamma_5 \, D_{x,x} \;\equiv\;  
            -\mbox{\rm tr} \,\gamma_5 \, (1 - \frac{1}{2\rho}D_{x,x})
   \label{eq:5}  
\end{equation}  
satisfies this requirement. 

The basic idea of Ref.~\cite{Hor04A} is to simultaneously monitor the physics (in this case 
topological susceptibility) and the geometry (in this case global behavior) as the space-time 
structure in $q(x)$ is built starting from the most intense point and then gradually lowering 
the intensity threshold for points included in the structure. Formally, we define the sets 
$\cS^q(f)$ (fraction supports) containing a fraction $f=N(\cS^q(f))/N(\Omega)$ of 
the most intense points as ranked by $|q(x)|$. Here $N(\Gamma)$ denotes the number of points 
contained in the subset $\Gamma$ of discretized torus $\Omega$. The fraction $f$ thus plays 
the role of the monitoring variable. We emphasize that what we mean by ``structure'' here 
is the partition of $\cS^q(f)$ into maximal path-connected~\footnote{By path-connected lattice 
set we mean a set of points any pair of which can be connected by a nearest-neighbor path 
passing only through points within the set.} subsets $\cS^q_i(f)$ containing the points with 
topological density of the same sign~\cite{Hor04A,Hor03A}. This definition of geometric 
structure emphasizes connectedness because extended connected concentrations of strong fields 
can naturally facilitate the long-distance physics (e.g. lead to inherently global Dirac 
eigenmodes~\cite{Hor04A} and Goldstone boson propagation). Considering sign coherence is 
physically motivated by the fact that quarks of preferred chirality are attracted to regions 
of definite sign.

To monitor the saturation of topological fluctuations, we compute the cumulative function 
$\cml(f)$ of topological susceptibility~\cite{Hor04A}, which represents the fraction of 
the total susceptibility carried by a fraction $f$ of the most intense points, i.e.
\begin{equation}
   \cml(f) \;\equiv\; \frac{\chi(f)}{\chi(1)}  \qquad\qquad
   \chi(f) \,\equiv\, \frac{\langle\, Q^2(f) \,\rangle \,-\, 
                            \langle\, Q(f)   \,\rangle^2}{V}
   \label{eq:7}
\end{equation}
where 
\begin{equation}
      Q(f) \,\equiv\, \sum_{x\in \cS^q(f)} q(x)     \qquad\qquad
      V \,\equiv\, a^4 N(\Omega)
      \label{eq:10}
\end{equation}
are the charge associated with the fraction support and the total volume respectively.
To monitor the global behavior in the structure, we compute two associated characteristic
functions. The first is the maximal linear size $\lsz_{max}(f)$ of the connected 
component, i.e. 
\begin{equation}
     \lsz_{max}(f) \;\equiv\; \max_{i}\, \{\, \lsz(\cS^q_i(f))\,\}
     \label{eq:14}
\end{equation}
Here the linear size $\lsz(\Gamma)$ for arbitrary $\Gamma \subset \Omega$ is the maximal
distance between two points of the set, i.e. 
$\lsz(\Gamma) \equiv \max \{\, |x-y| : \; x,y \in \Gamma \,\}$. The second characteristic 
is the point-wise average of the linear size over the points of the structure. More
precisely, to $x\in \cS^q_i(f)$ we assign the linear size 
$\lsz_{f}(x) \equiv \lsz(\cS^q_i(f))$ of the corresponding connected component, 
and take the average over $\cS^q(f)$
\begin{equation}
     \lsz_{pta}(f) \,\equiv\, \langle\, \lsz_{f}(x) \,\rangle_{\cS^q(f)}
     \label{eq:19}
\end{equation}
The motivation for introducing $\lsz_{max}(f)$ and $\lsz_{pta}(f)$ is that their 
values signal different regimes of global behavior in the structure. In particular, 
when $\lsz_{max}(f)$ reaches values $\lsz_{max}(f) \approx \lsz(\Omega)$ (close to 
the ``size of the space-time'') at some typical fraction $f_{max}$, then $f_{max}$ 
characterizes the onset of global behavior. In other words, an intensity threshold 
that corresponds to $f \ge f_{max}$ is low enough to include not only the most intense 
local concentrations of the charge distribution, but also the connecting ridges which 
join them into a global structure. In addition, for fractions $f \ge f_{max}$, where 
the largest connected piece is global, we would also like to know at what value of $f$ most 
of the points in the structure are part of the global component. The function $\lsz_{pta}(f)$ 
characterizes this property and its saturation at some value $f_{pta}\ge f_{max}$ indicates 
that global behavior is prevalent throughout the structure.

\begin{figure}[t]
\begin{center}
   \vskip -0.35in
   \centerline{
   \epsfxsize=15truecm\epsffile{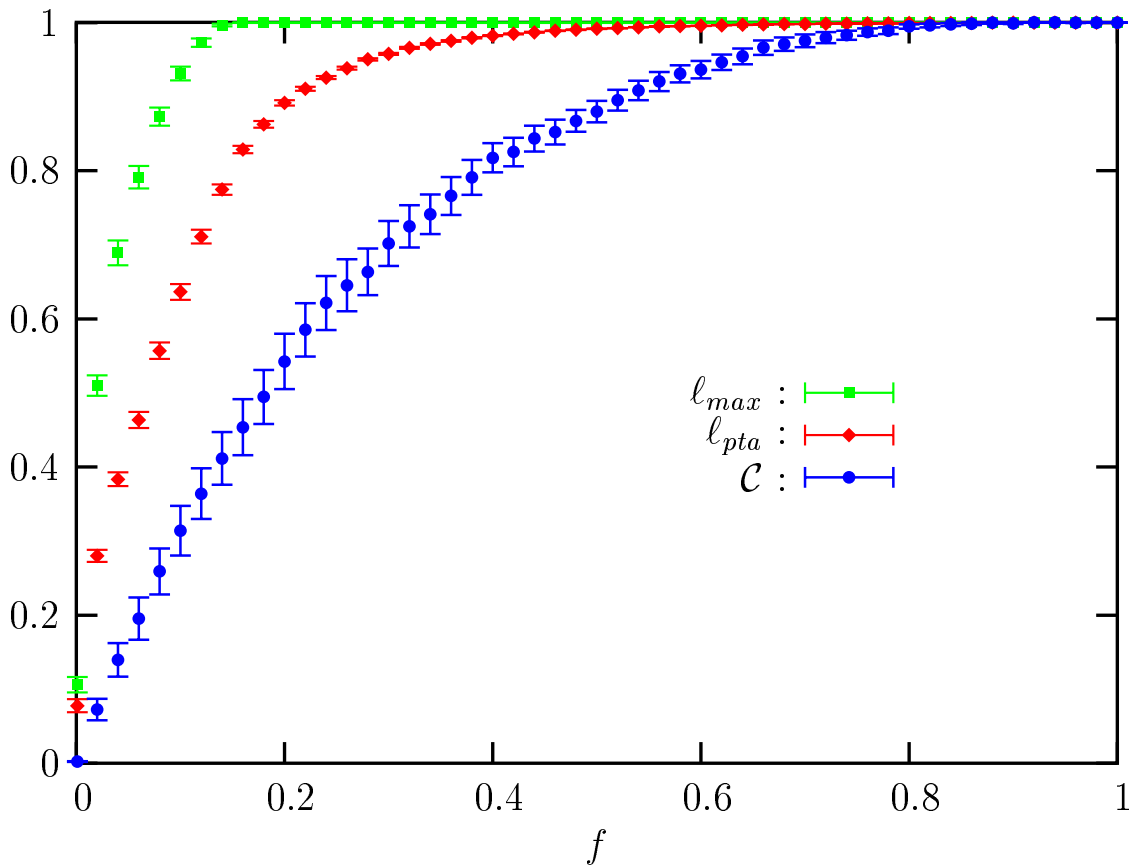}
   }
   \caption{The functions $\cml(f)$ (monitoring topological susceptibility) and
            $\lsz_{max}(f)$, $\lsz_{pta}(f)$ (monitoring the degree of global 
            behavior) for {\em full topological density} in ensemble with 
            renormalization group improved Iwasaki gauge action. Distances are
            measured in units of $\lsz(\Omega)$.} 
   \label{full_dens:fig}
   \vskip -0.4in
\end{center}
\end{figure}

We computed the above characteristics for the ensemble (51 independent configurations) 
of Iwasaki gauge action~\cite{Iwasaki} on an $8^4$ lattice with scale $a=0.165\,$fm 
determined from string tension. The physical volume of the system is thus $V=3$ fm$^4$. 
The topological charge density operator (\ref{eq:5}) with $\rho=1.368$ ($\kappa=0.19$) 
has been used to compute the site-by-site values of full topological charge density. 
The details of numerical implementation for overlap matrix--vector operation can be 
found in Ref.~\cite{Chen03}. The results are shown in Fig.~\ref{full_dens:fig}.

The data indicates that $\lsz_{max}(f)$ has a strict (full) saturation point at 
$f_{max}\approx 0.14$. Also, $\lsz_{pta}(f)$ exhibits a clear saturation ``knee''
at $f_{pta} \approx 0.20$. At the same time, the cumulative function of susceptibility
saturates more gradually and reaches a full saturation point (within the statistics) 
at $f_{\cml}\approx 0.80$. The central message of Fig.~\ref{full_dens:fig} is that 
at fractions where the geometric structure consists of localized pieces 
(i.e. $\lsz_{pta}(f) \le \lsz_{max}(f) \ll 1$)~\footnote{In what follows we will
always quote distances in units of $\lsz(\Omega)$, the maximal possible distance.}
only a small percentage of the total topological susceptibility is accounted for. 
In other words, topological susceptibility is not dominated by localized carriers 
of topological charge. Moreover, the susceptibility saturates only at fractions obviously 
larger than those at which the structure becomes prevalently global (i.e. $f>f_{pta}$). 
This result quantitatively justifies 
the proposition put forward in Ref.~\cite{Hor03A}, that the physically relevant
vacuum structure in topological charge density is {\em inherently global 
(super-long-distance)}. In other words, our data indicate that it is not correct
to view vacuum fluctuations of topological charge in terms of localized lumps 
of topological charge, while it strongly supports the idea that global extended objects 
should be used as a basis for valid physical analysis.

It should be noted here that our arguments started with the assumption that 
space-time regions containing the most intense fields (in this case topological
charge density) are most relevant for the physics in question (in this case 
topological susceptibility). This assumption is {\em a posteriori} justified 
by the fact that the computed $\cml(f)$ is a concave function on its domain 
within the errorbars. In other words, the ordering of space-time points by 
intensity results in monotonically decreasing contribution to susceptibility as 
the fraction is increased. 
\medskip

\noindent
{\bf 3. Low Energy.} 
Our analysis in the previous section was carried out for full topological charge
density, i.e. with fluctuations at all scales up to the lattice cutoff included. 
An attractive feature of topological charge density operators based on 
Ginsparg-Wilson fermionic actions is the possibility of defining an 
{\em effective topological field}~\cite{Hor02B} via the eigenmode expansion of 
Eq.~(\ref{eq:5}), i.e.
\begin{equation}
   q^{\Lambda^F}(x) \;=\; 
   -\sum_{\lambda\le\Lambda^F a} (1 - \frac{\lambda}{2\rho})\, c^{\lambda}(x) 
   \label{eq:24}  
\end{equation}
where $c^\lambda(x) = \psi^{\lambda \,\dagger}(x) \gamma_5 \psi^{\lambda}(x)$ 
is the local chirality of the mode $\psi^{\lambda}$ with eigenvalue $\lambda$. 
The effective field $q^{\Lambda^F}(x)$ describes topological fluctuations up to 
energy scale $\Lambda^F$ defined by the probing chiral fermion. Since zero modes 
are included in the expansion for arbitrary $\Lambda^F$, we have
$\sum_x q(x) = Q = \sum_x q^{\Lambda^F}(x) = n_{-}-n_{+}$. Here $n_{-(+)}$
is the number of zero modes with negative (positive) chirality. This has three
noteworthy implications~\cite{Hor02B}. (a) The topological susceptibility is 
independent of $\Lambda^F$. (b) Similarly to $q(x)$, the effective density is 
a perfect topological field in the sense that (for generic backgrounds) its 
global sum is strictly stable with respect to small local changes of the gauge 
field. (c) $q^{\Lambda^F}(x)$ satisfies the index theorem on the lattice in 
the same sense that $q(x)$ does~\cite{Has98A}.

\begin{table}[b]
  \centering
  \begin{tabular}{cccccc}
  \hline\hline\\[-0.4cm]
  \multicolumn{1}{c}{ensemble}  &
  \multicolumn{1}{c}{$\beta$}  &
  \multicolumn{1}{c}{$a$ [fm]}  &
  \multicolumn{1}{c}{$V_{lat}$}  &
  \multicolumn{1}{c}{$\quad$configs$\quad$} &
  \multicolumn{1}{c}{$\;$eigen-pairs$\;$} \\[2pt]
  \hline\\[-0.4cm]
   $\;\cE_1\;$ & $\quad 5.91 \quad$ & $\quad 0.110 \quad$ & $\quad 12^4 \quad$ 
           & $\quad 20 \quad$ & $\quad 12 \quad$\\
   $\;\cE_2\;$ & $\quad 6.07 \quad$ & $\quad 0.082 \quad$ & $\quad 16^4 \quad$ 
           & $\quad 20 \quad$ & $\quad 16 \quad$\\
   $\;\cE_3\;$ & $\quad 6.37 \quad$ & $\quad 0.055 \quad$ & $\quad 24^4 \quad$ 
           & $\quad 20 \quad$ & $\quad 12 \quad$\\
\hline \hline
\end{tabular}
\caption{Ensembles of Wilson gauge configurations. Zero modes and fixed
number of lowest complex eigen-pairs were computed for each configuration.}
\label{ensemb_tab} 
\end{table}

   \begin{figure}
   \begin{center}
     \vskip -0.20in
     \centerline{
     \epsfxsize=8.2truecm\epsffile{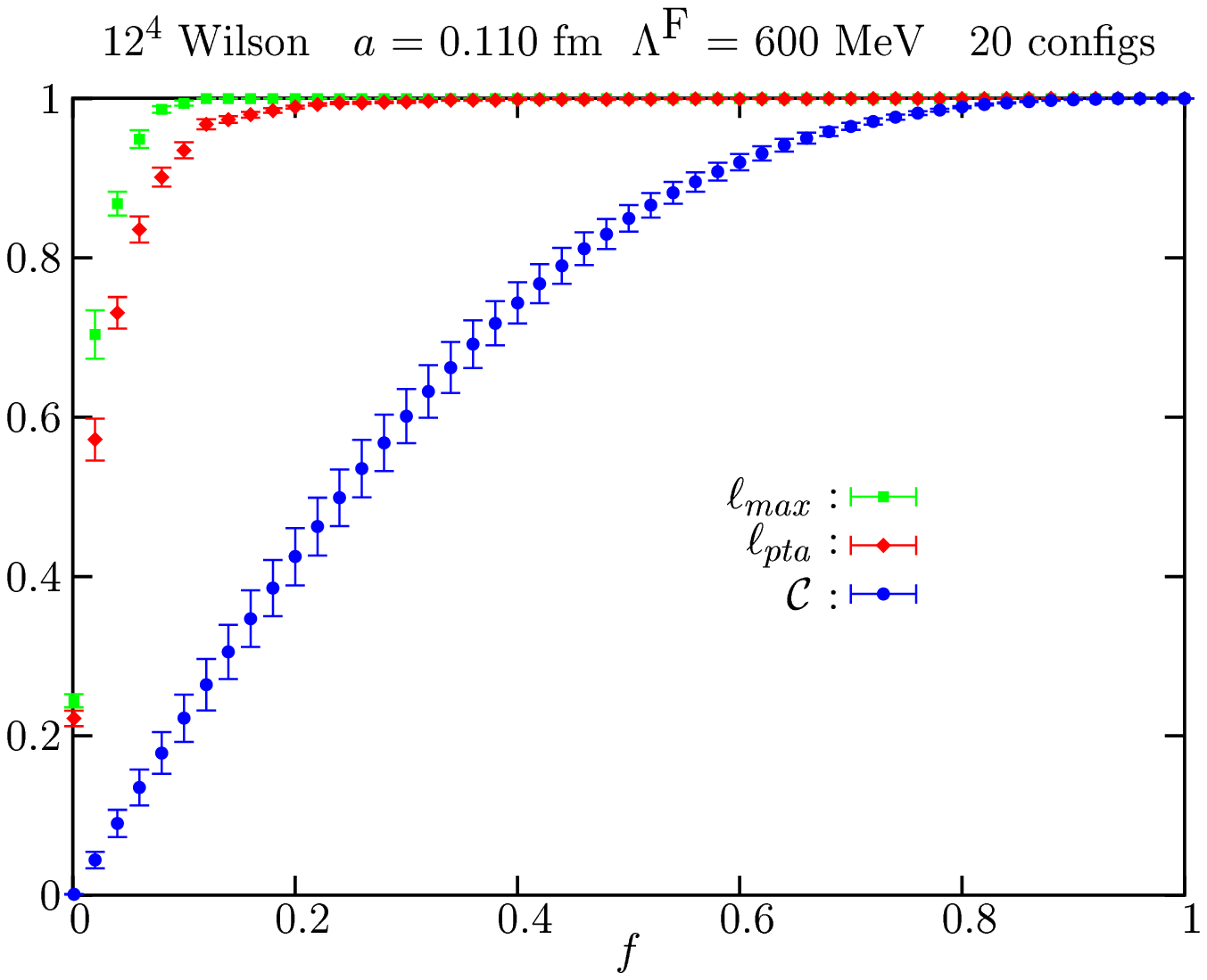}
     \hskip -0.00in
     \epsfxsize=8.2truecm\epsffile{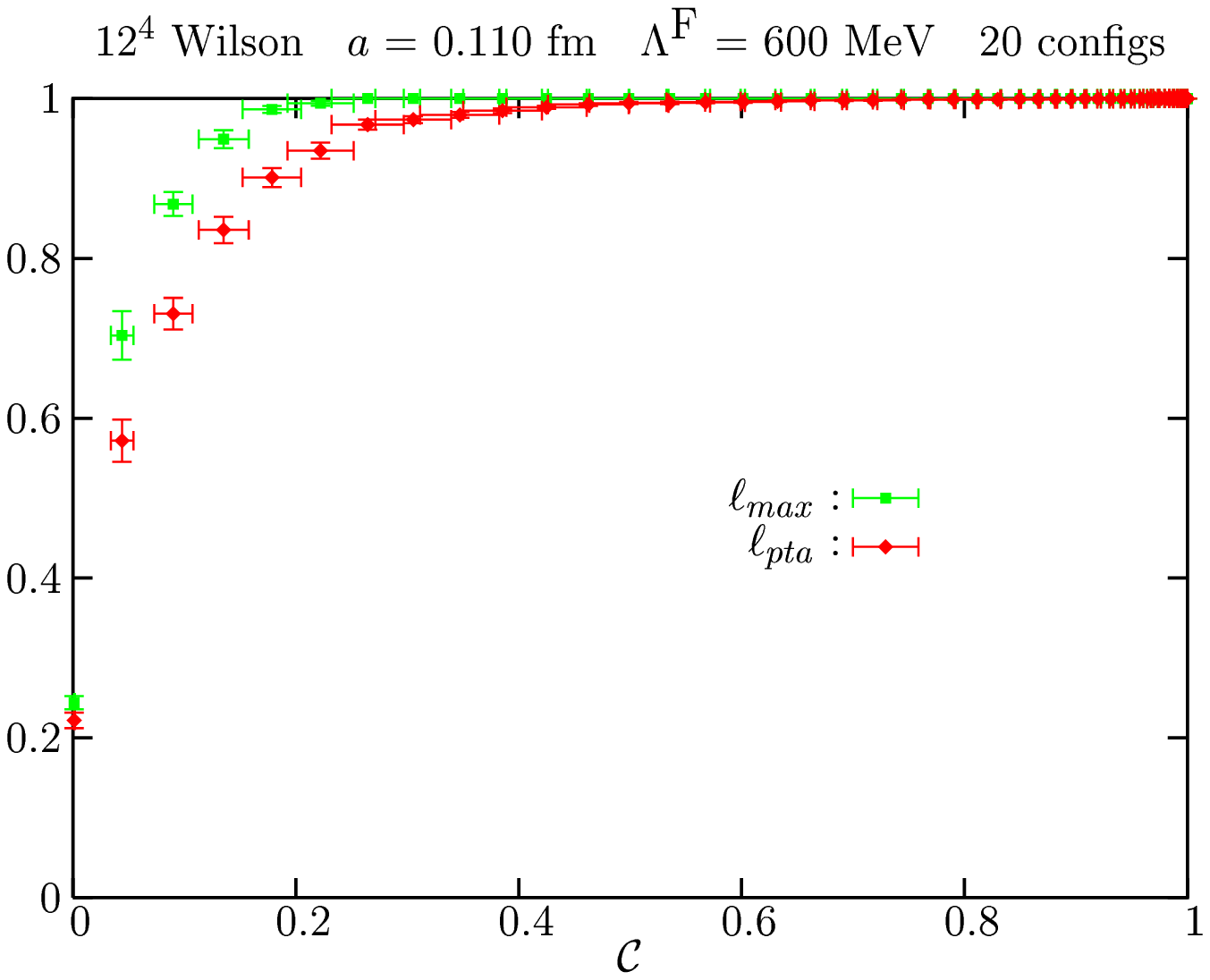}
     }
     \vskip 0.10in
     \centerline{
     \epsfxsize=8.2truecm\epsffile{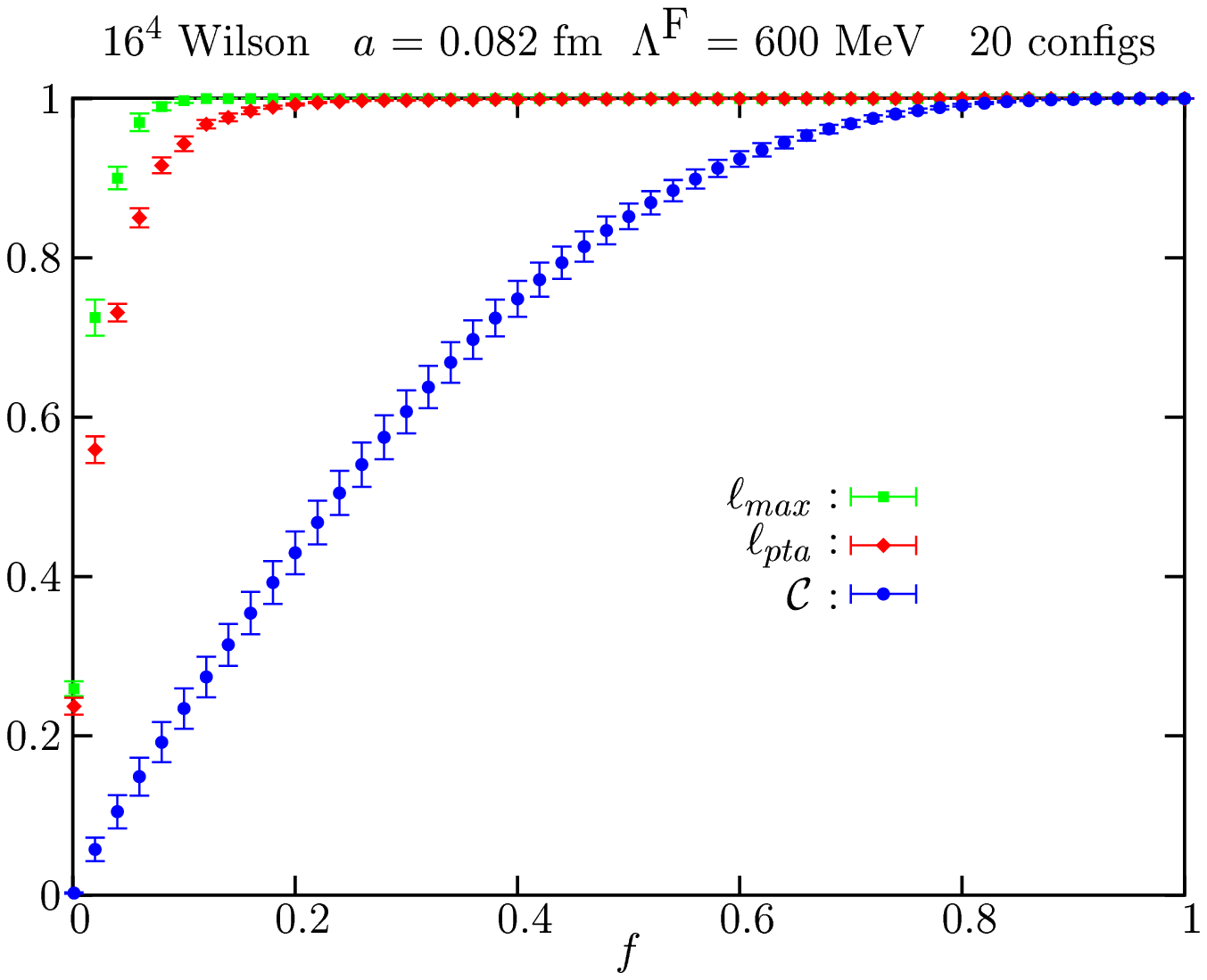}
     \hskip -0.00in
     \epsfxsize=8.2truecm\epsffile{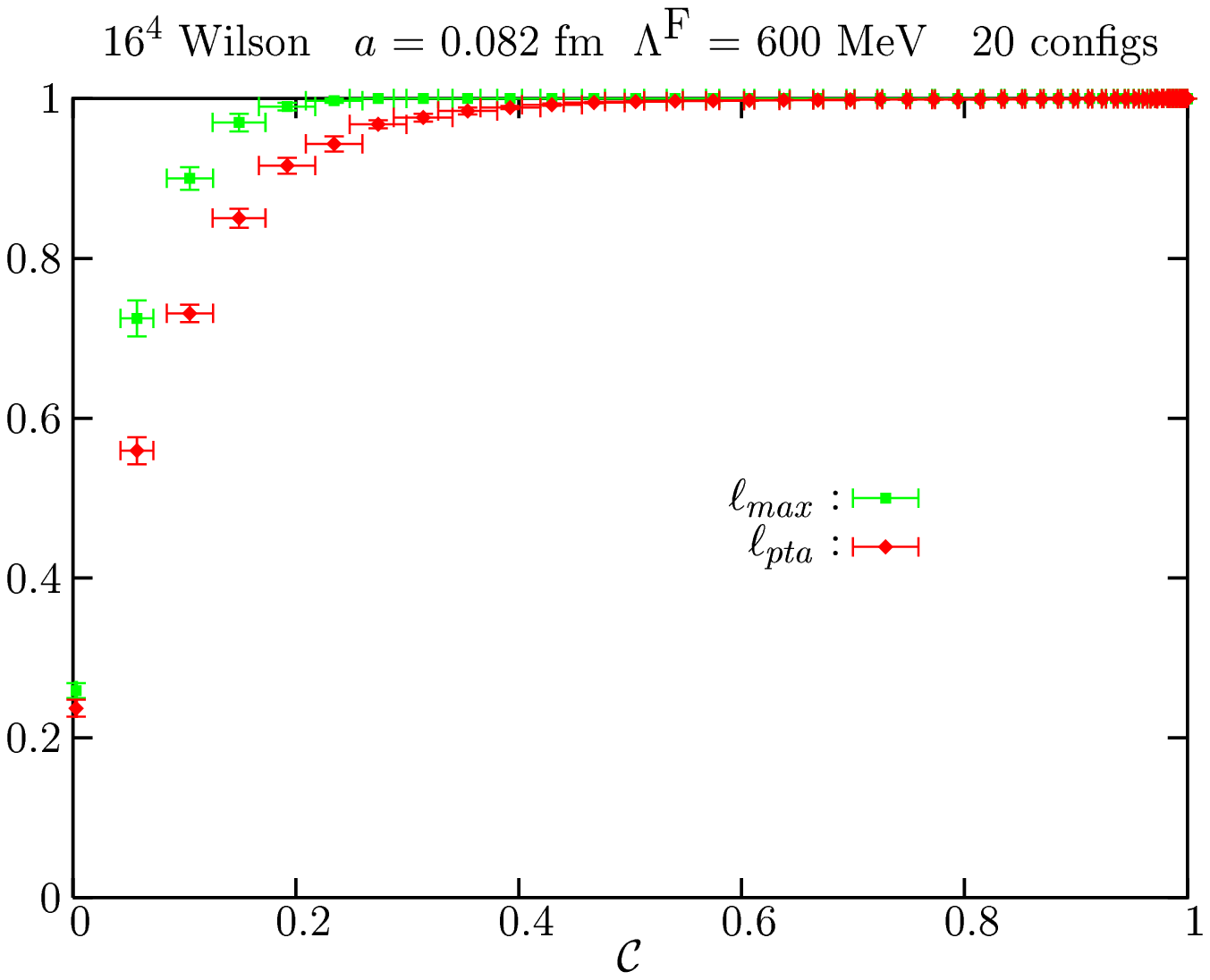}
     }
     \vskip 0.10in
     \centerline{
     \epsfxsize=8.2truecm\epsffile{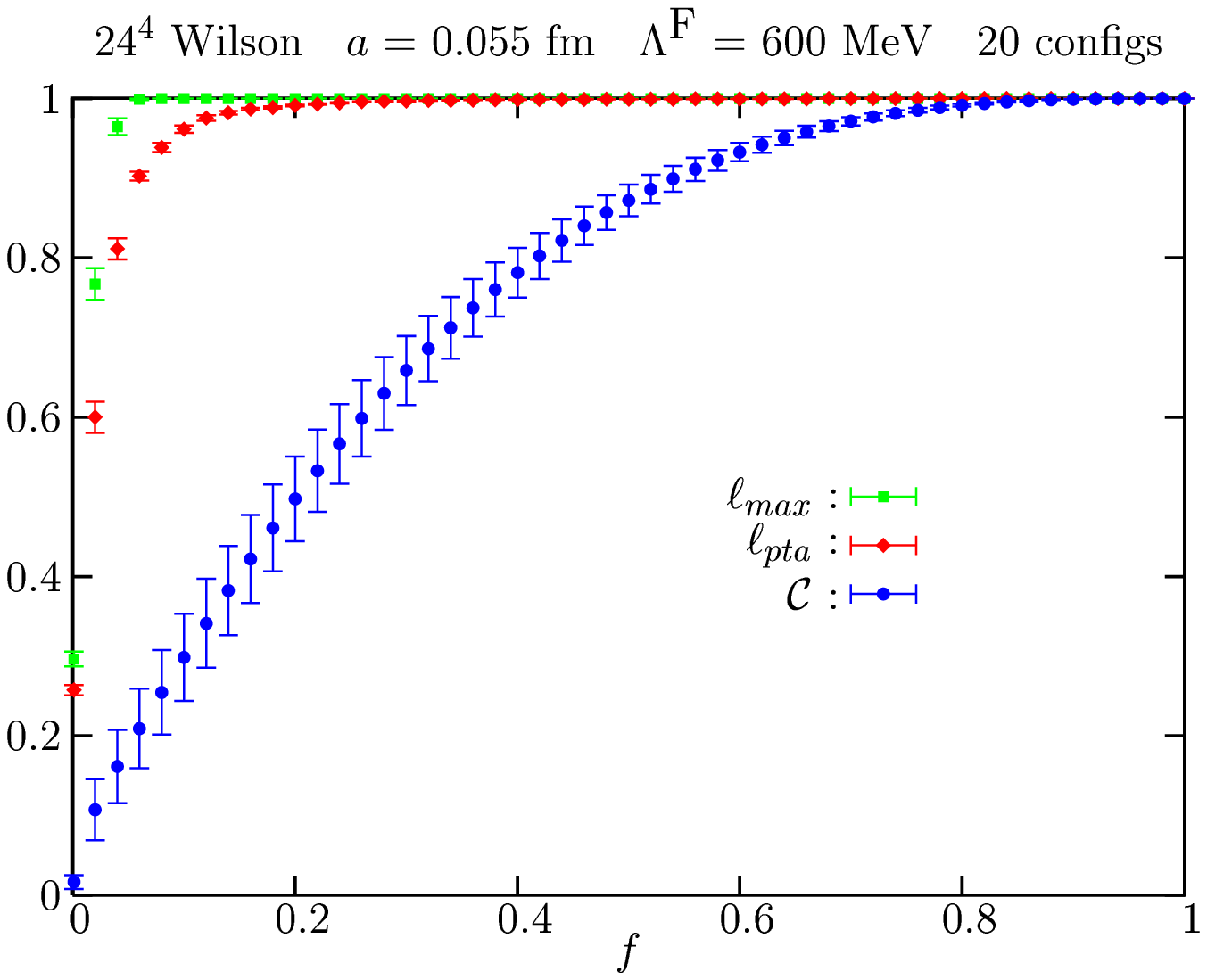}
     \hskip -0.00in
     \epsfxsize=8.2truecm\epsffile{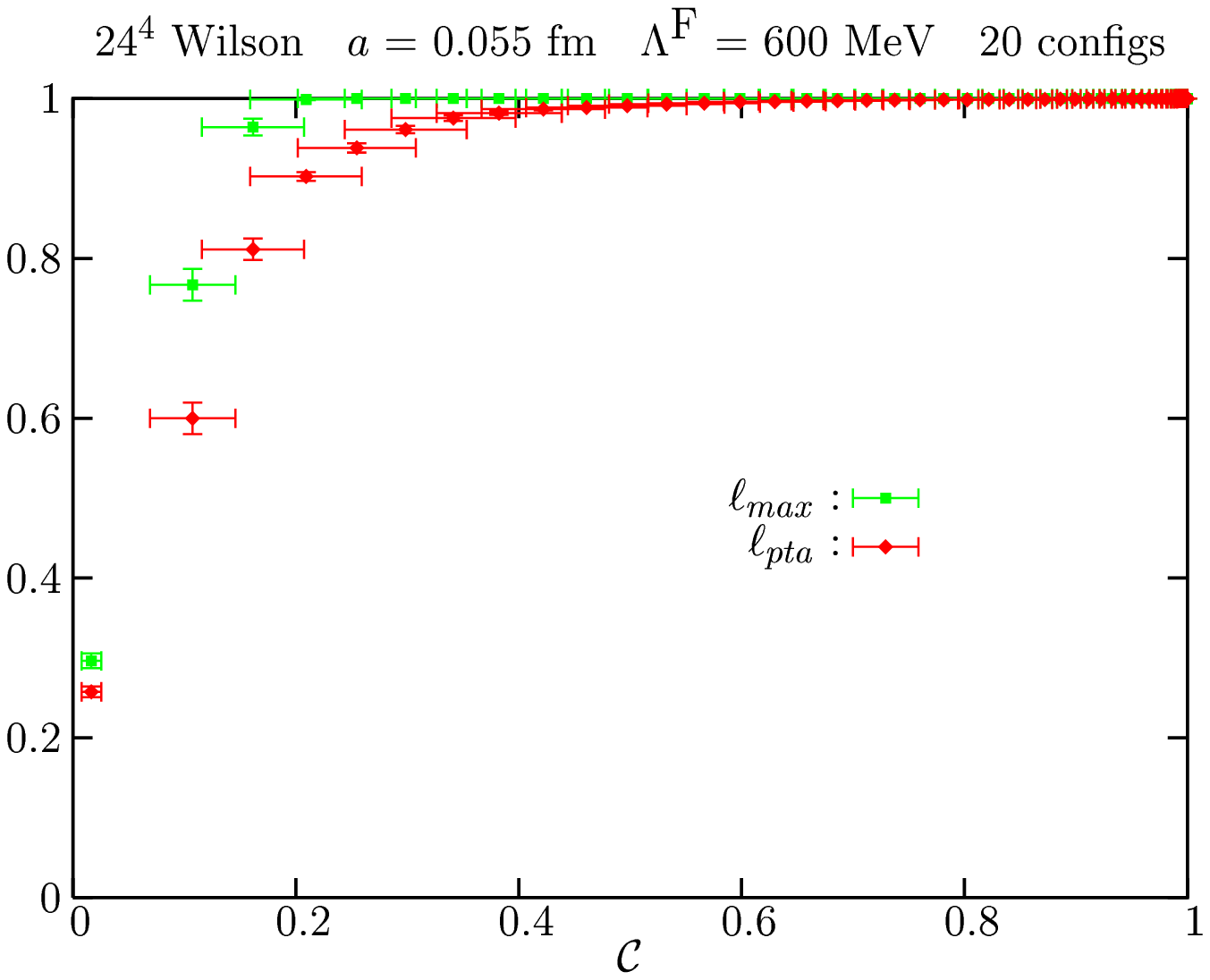}
     }
     \caption{The cumulative function of topological susceptibility and the global 
              characteristics of the geometric structure at low energy. See the 
              discussion in the text. Distances are measured in units of 
              $\lsz(\Omega)$.}
     \label{eff_dens:fig}
   \end{center}
   \end{figure}

The introduction of the effective topological density opens the possibility of 
describing the space-time structure of topological fluctuations in 
a scale-dependent manner as is appropriate in field theory~\cite{Hor02B}. 
The emerging precise relation between the structure at low energy and 
the fundamental structure will be described elsewhere. Here we focus on 
the issue of global behavior at low energy and how it compares to the behavior
at the fundamental level. To do so, we have constructed effective densities
at $\Lambda^F=600$ MeV for three ensembles of Wilson gauge action with 
parameters described in Table~\ref{ensemb_tab}. Lattices at different cutoffs 
have the same physical volumes $V \approx 3$ fm$^4$. The quoted values of 
lattice spacing were obtained via the Sommer parameter using the interpolation 
formula given in Ref.~\cite{Sommer}. For each configuration in a given ensemble 
we have calculated the zero modes and a fixed number of complex-conjugate
eigenmode pairs specified in Table~\ref{ensemb_tab}. The Ritz variational 
method~\cite{Ritz} was used for this calculation. A typical accuracy of calculated 
eigenvalues (as measured by differences of complex-conjugate pairs) was one 
part in $10^5$ or better. The largest cutoff at which the effective density 
could be constructed for $\cE_3$ was just above $\Lambda^F=600$ MeV (larger 
for $\cE_1$ and $\cE_2$).

We have calculated the cumulative function of topological susceptibility
$\cml(f)$ and the global characteristics $\lsz_{max}(f)$ and $\lsz_{pta}(f)$
for the three ensembles with the results shown in Fig.~\ref{eff_dens:fig}
(left column). One can see that the main change at low energy compared
to the fundamental structure is that the functions $\lsz_{max}(f)$ and 
$\lsz_{pta}(f)$ saturate even faster. For example, $\lsz_{max}(f)$ in 
$\cE_3$ ($a=0.055$ fm) reaches full saturation point at $f_{max} < 0.06$. 
We thus conclude that the {\em inherently global} nature of the topological 
structure in the QCD vacuum is even more apparent at low energy. Moreover, 
the wide range of cutoffs spanned by our data shows convincingly that 
this conclusion will not change in the continuum limit.

To see the relation between the degree of saturation in susceptibility
and in global behavior more directly, we follow Ref.~\cite{Hor04A} and
eliminate the ``monitoring variable'' $f$ in favor of $\cml$. This defines
the functions $\lsz_{max}(\cml)$ and $\lsz_{pta}(\cml)$ that quantify
the degree of global behavior for any given value of susceptibility 
saturation. These functions are plotted in the right column of 
Fig.~\ref{eff_dens:fig} and exhibit the strongly concave behavior 
characteristic of inherently global structure. From the data one can see 
that the structure becomes prevalently global when it carries only 
$\cml \approx 0.2$ of total susceptibility.  

\begin{figure}[t]
\begin{center}
   \vskip -0.35in
   \centerline{
   \epsfxsize=15truecm\epsffile{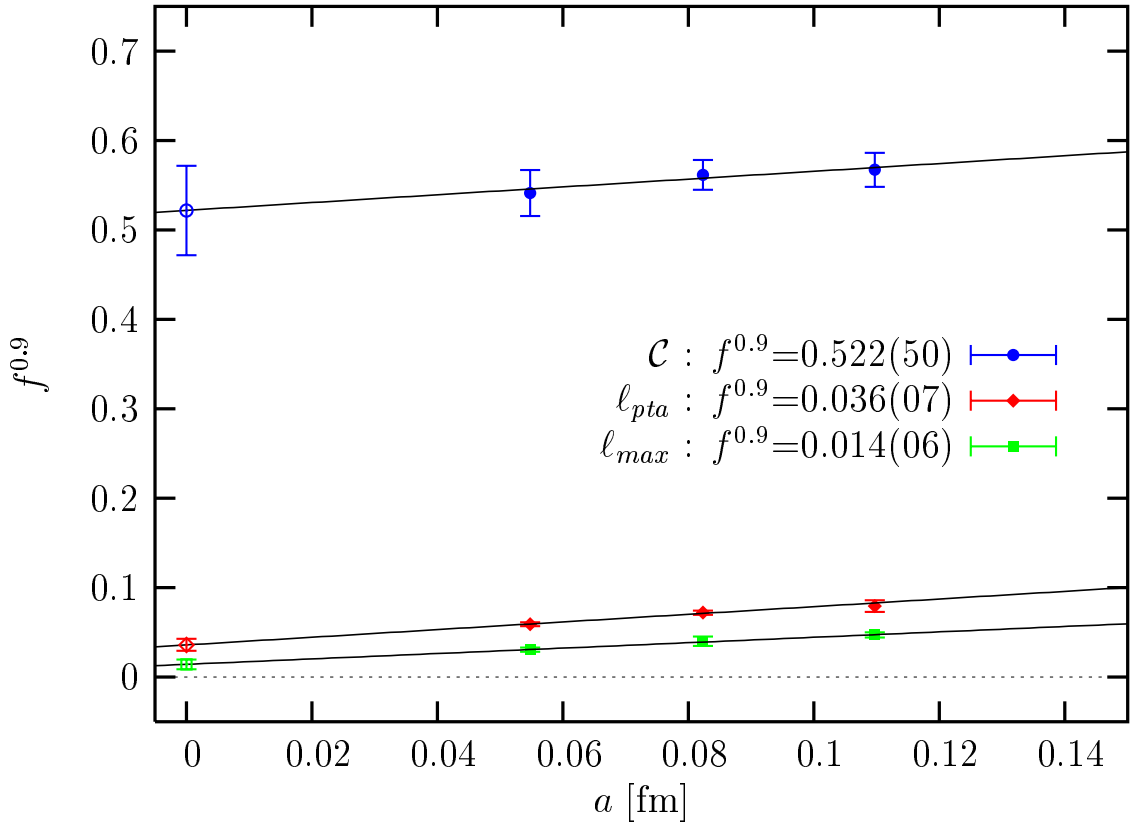}
   }
   \caption{The continuum extrapolation of $f^{0.9}$ associated with
            $\lsz_{max}(f)$, $\lsz_{pta}(f)$ and $\cml(f)$. The linearly 
            extrapolated values are quoted in the plot.} 
   \label{f09:fig}
   \vskip -0.4in
\end{center}
\end{figure}

To quantify the vast discrepancy between the fractions of space-time volume
characterizing the appearance of global behavior and the saturation of
susceptibility, we have calculated the scale dependence of fractions 
at which $\lsz_{max}(f)$, $\lsz_{pta}(f)$ and $\cml(f)$ reach the reference 
saturation value $y(f^{0.9})=0.9$. In Fig.~\ref{f09:fig} we plot $f^{0.9}(a)$ 
for the three characteristic functions. In all three cases we performed a linear 
extrapolation to the continuum limit with the extrapolated results quoted 
in the plot. This calculation shows that 90\% saturation of susceptibility 
requires inclusion of about 50\% of space-time occupied by most intense 
topological charge density in the continuum limit. At the same time, 
the 90\% saturation of $\lsz_{max}(f)$ and $\lsz_{pta}(f)$ requires less 
than 2\% and 4\% of space-time respectively. This represents a rather striking 
demonstration that topological structure in QCD vacuum is inherently global 
(super-long-distance) at low energy.

{\bf 4. Conclusions.}
The goal of the study described here was to justify the proposition of 
Ref.~\cite{Hor03A} that the ordered vacuum structure in topological charge density,
first observed there, is based on underlying {\em global} connected object(s) 
present in the vacuum. The original suggestion of Ref.~\cite{Hor03A} was based
on the purely geometrical observation that the regions of space-time containing
the most intense topological density have a strong preference to organize into a global 
long-range structure via the formation of a {\em skeleton}~\footnote{In the language 
developed in Ref.~\cite{Hor04A} the formation of a skeleton corresponds to saturation
of $\lsz_{pta}$. This definition is simpler than the one used in Ref.~\cite{Hor03A} 
and leads to similar quantitative results.}, i.e. the minimal 
hard-core substructure exhibiting the prevalently global behavior. In order 
to ascertain that a description of physics entails the existence of global geometric 
objects rather than some hypothetical localized objects forming the skeleton, it is 
necessary to evaluate the contribution of such objects to the relevant physical 
observable. Ref.~\cite{Hor04A} developed a formalism for performing such a calculation 
in the case of topological susceptibility. We adopted this formalism and the results 
presented here demonstrate that the global behavior dominates at decisively smaller 
fractions of space-time than those that are necessary to saturate the topological 
susceptibility. 
This is true both at the fundamental level and at low energy where the difference 
of relevant fractions is strikingly large. From these observations we conclude that 
the vacuum topological structure behaves as {\em inherently global (super-long-distance)} 
in QCD.~\footnote{We note that the low-lying eigenmodes of overlap Dirac operator are 
also concentrated on inherently global objects as pointed out in Ref.~\cite{Hor02A}
and demonstrated in Ref.~\cite{Hor04A}.}
The relevance of this result resides in the fact that it directly indicates that 
a valid theoretical description of topological charge fluctuations in QCD vacuum needs 
to be based on global brane-like objects.

\bigskip\medskip
\noindent
{\bf Acknowledgments:} 
This work was supported in part by U.S. Department of Energy under grants 
DE-FG05-84ER40154 and DE-FG02-95ER40907. I.H. acknowledges the hospitality of Kavli 
Institute of Theoretical Physics (and the organizers of the program ``Modern 
Challenges for Lattice Gauge Theory'') where this manuscript was finalized.

\end{document}
\bye